\documentclass[onecolumn,amsfonts,amssymb,amsmath,floatfix]{article}
\usepackage[dvips]{graphicx}
\usepackage{natbib}
\usepackage{bm}
\bibpunct{(}{)}{;}{a}{}{,}

\title{Scalar potential model of the Pioneer Anomaly}

\author{J.C. Hodge$^{1}$\thanks{E-mail:jch9496@blueridge.edu}\thanks{Visiting from XZD Corp., 16 Hosta Ln., Brevard, NC, 28712, E-mail:{scjh@citcom.net}}\\
$^{1}${Blue Ridge Community College, 100 College Dr., Flat Rock, NC, 28731-1690}}

\begin{document}

\maketitle

\begin{abstract}
The unexplained sunward acceleration $a_\mathrm{P}$ of the Pioneer 10 (P10) and the Pioneer 11 (P11) spacecraft remains a mystery.  A scalar potential model (SPM) that derived from considerations of galaxy clusters, of redshift, and of H{\scriptsize{I}} rotation curves of spiral galaxies is applied to the Pioneer Anomaly.  Matter is posited to warp the scalar potential $\rho$ field.  The changing $\rho$ field along the light path causes the Pioneer Anomaly.  The SPM is consistent with the general value of $a_\mathrm{P}$; with the annual periodicity; with the differing $a_\mathrm{P}$ between the spacecraft; with the discrepancy between {\textit{Sigma}} and CHASMP programs at P10 (I) and their closer agreement at P10 (III); with the slowly declining $a_\mathrm{P}$; with the low value of $a_\mathrm{P}$ immediately before the P11's Saturn encounter; with the high uncertainty in the value of $a_\mathrm{P}$ obtained during and after the P11's Saturn encounter; and with the cosmological connection suggested by $a_\mathrm{P} \approx cH_\mathrm{o}$.  The effect of the $\rho$ field warp appears as the curvature of space proposed by general relativity (GR).  The Hubble Law and $a_\mathrm{P} \approx cH_\mathrm{o}$ are manifestations of the Newtonian spherical property.  Therefore, gravitational attraction, the equivalence principle, and the planet ephemeris remain as described by GR.  GR corresponds to the SPM in the limit in which the Sources and Sinks may be replaced by a flat and static $\rho$ field such as between cluster cells and on the Solar System scale at a relatively large distance from a Source or Sink.
\end{abstract}

Pioneer Anomaly -- Gravity Tests -- cosmology:theory 
PACS 04.80.Cc, 95.30.Sf, 98.80.-h 

\section{INTRODUCTION}

That an unexplained blueshift exists in the electromagnetic (EM) radio signal from the Pioneer 10 (P10) and Pioneer 11 (P11) (PA) is well established \citep{ande02,toth}.  Several models have been proposed to explain the PA \citep{ande02}.  A currently popular interpretation of the PA is that the Pioneer spacecraft are being subjected to a force that causes a sunward acceleration $a_\mathrm{P} \approx (8.74 \pm 1.33) \times 10^{-8}$ cm s$^{-2}$.  That $a_\mathrm{P} \approx c H_\mathrm{o}$, where $c$ (cm s$^{-1}$) is the speed of light and $H_\mathrm{o}$ (s$^{-1}$) is the Hubble constant, suggest a cosmological connection to PA.  However, the PA exceeds by at least two orders of magnitude the general relativity (GR) corrections to Newtonian motion.

The PA is experimentally observed as a frequency shift but expressed as an apparent acceleration.  The PA could be an effect other than a real acceleration such as a time acceleration \citep{ande02,niet05a} or an effect of an unmodeled effect on the radio signals.  Although unlikely, a currently unknown systematics effect is not entirely ruled out.

Data for the existence of $a_\mathrm{P}$ from the Galileo, Ulysses, Voyager, and Cassini spacecraft are inconclusive \citep{ande02,niet05b}. 

In addition to the Sun directed blueshift, there are other characteristics of the PA \citep{ande02}.  The PA has an apparent annual periodicity with an amplitude of approximately 1.6$\times 10^{-8}$ cm s$^{-2}$.  Although within uncertainty limits, the P11 anomaly $a_\mathrm{P11} \approx 8.55 \times 10^{-8}$ cm s$^{-2}$ may be slightly larger than the P10 anomaly $a_\mathrm{P10} \approx 8.47 \times 10^{-8}$ cm s$^{-2}$ \citep[Section V.A., Section VI.A., Section VI.D., and Section VIII.G.]{ande02}.  The $a_\mathrm{P}$ calculation by the {\textit{Sigma}} and CHASMP program methods for P10 (I) and P10 (II) show a discrepancy while showing consistency for P10 (III) \citep[Table I]{ande02}.  The $a_\mathrm{P}$ of both spacecraft may be declining with distance \citep[as shown by the envelope in Fig.~1]{tury}.  The blue shift of the PA is significantly smaller \citep{niet05a} immediately before P11's Saturn encounter.  The value of $a_\mathrm{P}$ averaged over a period during and after the Saturn encounter had a relatively high uncertainty \citep{niet05a}.  

An obstacle to a new gravitational physics explanation of the PA is that a modification of gravity large enough to explain the PA is in contradiction to planetary ephemeredes unless the Equivalence principle is violated \citep{ande02}.  The common opinion is that cosmic dynamics according to GR has far too little influence in galaxies to be measurable and that the expansion of the universe is essentially negligible for scales up to galactic clusters \citep{coop,sell}.  Further, the expansion of the universe indicated by redshift $z$ has a sign opposite to $a_\mathrm{P}$.  Several new physics models have been proposed \citep{ande02,bert}.  \citet{bert} concluded a scalar field is able to explain the PA.

A scalar potential model (SPM) was derived from considerations of galaxy clusters \citep{hodg}.  The SPM posits a scalar potential $\rho$ field exists.  The Sources of the $\rho$ field are in the center of spiral galaxies.  Sinks are in elliptical and other galaxies.  The $\rho$ field was posited to flow from Sources to Sinks like heat or fluid.  The SPM proposed the $\rho$ field caused a non-Doppler redshift or blueshift of photons traveling through it.  The resulting equation was applied to $z$ and discrete redshift of galaxies.  

This Paper argues that matter causes a static, non-flowing warp of the $\rho$ field that causes the PA.  The $\rho \propto -R^{-1}$ of the warp induces the $H_\mathrm{o}$ value and the connection to $z$ observations.  That is, the PA is an unaccounted effect on only the EM signal.  Therefore, gravitational attraction, the equivalence principle, and the planet ephemeris remain as described by GR.  

In section~\ref{sec:model}, the SPM is described and an equation to calculate $a_\mathrm{P}$ is derived.  The derived equation is used to calculate $a_\mathrm{P}$ in Section~\ref{sec:results}.  The discussion and conclusion are in Section~\ref{sec:disc}.

\section{\label{sec:model}Model}
In addition to the propositions of the SPM, matter is posited to cause a static\footnote{``Static'' such as caused by a stationary electron in a stationary EM field.  Sources induce a flow such as a source of heat or fluid.} warp in the $\rho$ field in accordance with the Newtonian spherical property.  Because the $\rho$ field near matter must attract other matter, the matter decreases the $\rho$ field.  The $\rho$ field then causes matter attraction according to established gravitational physics and causes the frequency change of the EM signal.  ``Static'' because matter is neither a Source nor a Sink of energy.  Matter merely modifies the energy flowing from Sources to Sinks.  The PA is an effect on only the EM signal and is a blueshift superimposed on the Doppler redshift of the receding spacecraft.

The amount of warp $\rho_\mathrm{m}$ at a point in space was posited to be equal to $GM/R$, where $G$ is the Newtonian gravitational constant, $M$ is the mass of a body, and $R$ is the distance from the center of mass to the point where $\rho$ is calculated.  That is,
\begin{equation}
\rho_\mathrm{m}=- \sum_i^N GM_i/R_i 
\label{eq:1},
\end{equation}
where N is the number of bodies used in the calculation.

The $K_\mathrm{min}$ term in the equation derived by \citet{hodg} resulted from the flow from Sources.  The $K_\mathrm{vp}$ term results from the relative movement of galaxies.  Therefore, $K_\mathrm{min} =0$ and $K_\mathrm{vp} = 0$ for the static warp field of matter in the Solar System.  Because the $K$ factors were calculated in a flowing $\rho$ field, the static warp requires other values of the $K$ factors.  The resulting equation for the calculated redshift $z_\mathrm{c}$ is
\begin{equation}
z_\mathrm{c} = \mathrm{e}^{-X} - 1
\label{eq:26},
\end{equation}
where
\begin{equation}
X= K_\mathrm{dp} D_\mathrm{l} P + K_\mathrm{p} P + K_\mathrm{f} F 
\label{eq:27},
\end{equation}
the terms are defined in \citet{hodg}, $ D_\mathrm{l}=2D$ (AU) is the distance the radio signal travels, and $D$ (AU) is the geocentric distance to the spacecraft.  The $P$ is a measure of the amount of $\rho$ the EM signal travels through. The $F$ is a measure of the inhomogeneity (turbulence) of $\rho$ the EM signal travels through. 

\section{\label{sec:results}Results}

\subsection{\label{sec:sample}Sample}

The mass of the Kuiper belt of approximately 0.3 Earth masses \citep{tepl} and the asteroid belt of approximately one Earth mass were included in the mass of the Sun.  The ephemeris including $GM$ of the Sun, planets, dwarf planets, and moons of Saturn were obtained from JPL's Horizon web site\footnote{http://ssd.jpl.nasa.gov/horizons.cgi} in November and December 2006.  The planets barycenter data were used for the calculation except for the Earth and its moon and except when considering the Saturn encounter of P11.  When considering the Saturn encounter, the $GM$ of the moons of Saturn without $GM$ data in the Horizon web site were calculated from the relative volume and mass of the other moons of Saturn.  The data were taken from the ephemeris for 00$^h$00$^m$ of the date listed except for the Saturn encounter where hourly data were used.

The $a_\mathrm{P}$ data were obtained from Table 2 of \citet{niet05a}.  The calculation of $\rho$ starting from the surface of the Earth along the line-of-sight (LOS) to the position of the spacecraft and back used a Visual Basic program.  Note the calculation of $F$ is direction dependent.  The Galaxy's effective mass was calculated from the revolution of the Sun about the center of the Galaxy and, for simplicity, assumed a spherically symmetric galactic matter distribution.  For the calculations, the Galaxy center of mass was positioned at 8 kpc from the Sun in the direction of Sgr A*.  The $\rho$ field due to the Source was assumed to be flat across the solar system.  Therefore, the effective mass at the center of the galaxy accounts for both the variation of $\rho$ from the Source and the true mass within the Galaxy \citep{hodg2}. 

Equation~(\ref{eq:26}) was used to calculate the $z_\mathrm{c}$ for each spacecraft on each date.  The calculated PA acceleration $a_\mathrm{Pc}$ (cm s$^{-2}$) \citep{ande02} is 
\begin{equation}
a_\mathrm{Pc}= -c^2 z_\mathrm{c}/D_\mathrm{l}
\label{eq:3}.
\end{equation}

\begin{figure*}
\begin{minipage}[t]{0.45\textwidth}
\includegraphics[width=\textwidth]{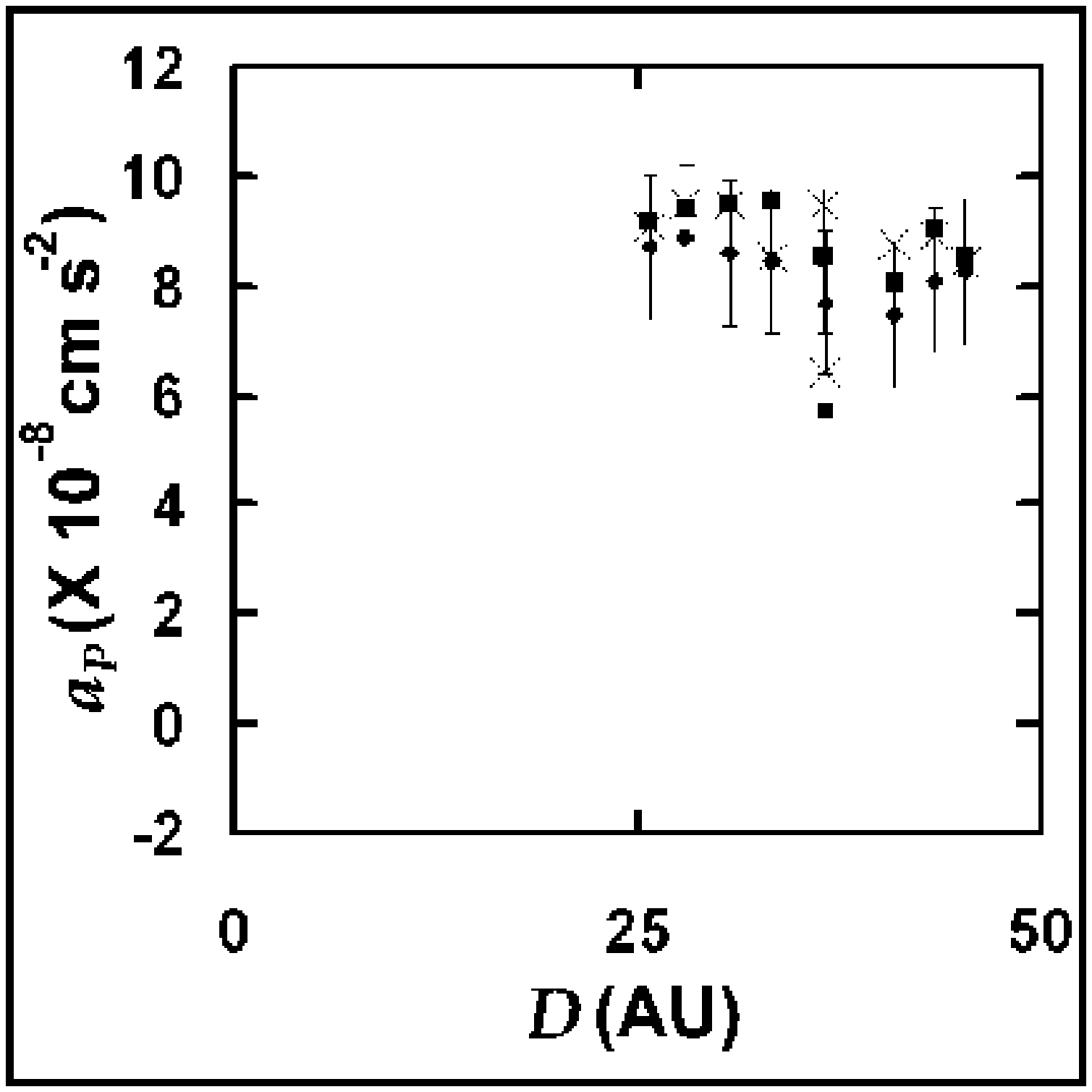}
\end{minipage}
\hspace{0.1in}
\begin{minipage}[t]{0.45\textwidth}
\includegraphics[width=\textwidth]{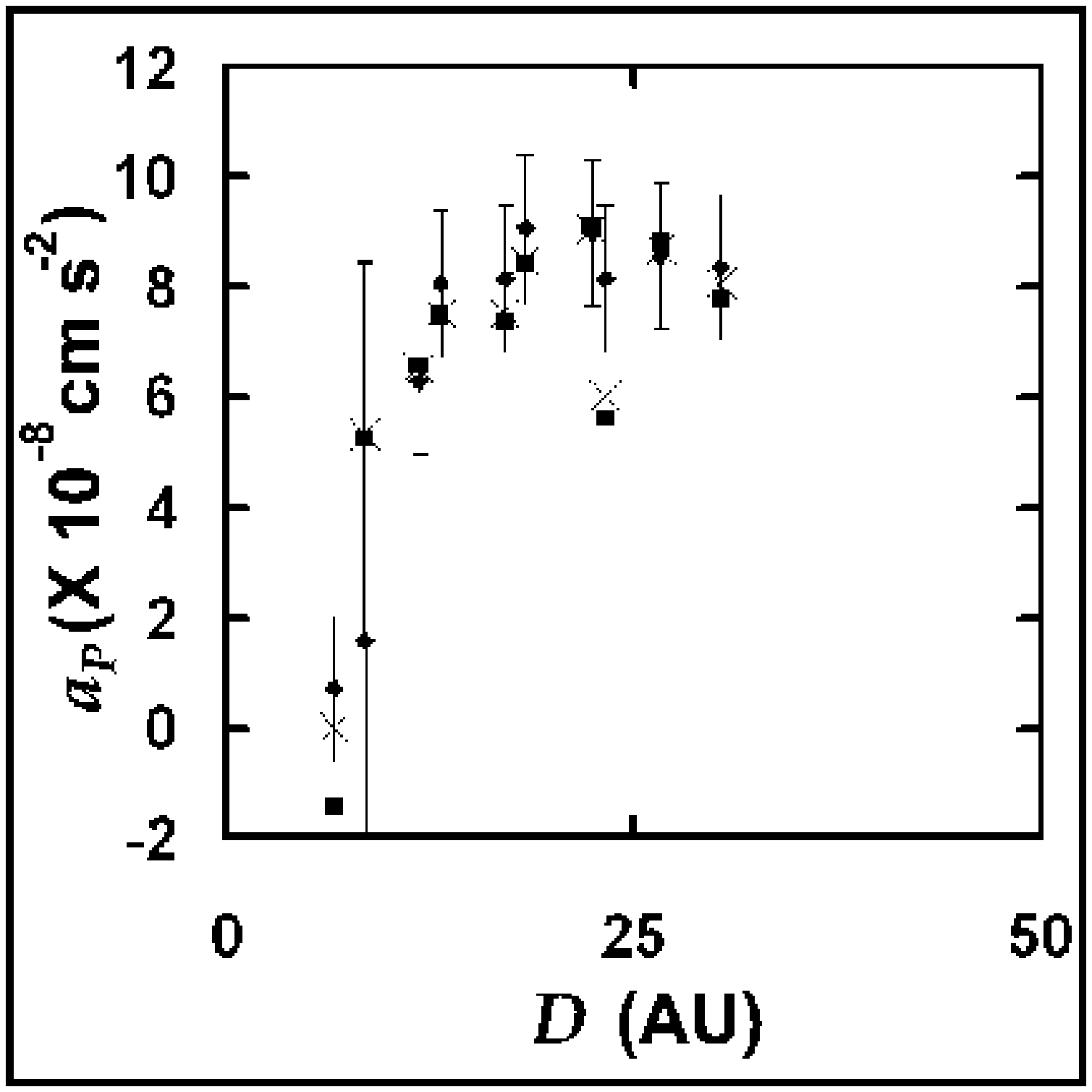}
\end{minipage}
\caption{Plots of the $a_\mathrm{P}$ data versus geocentric distance $D$ for P10 (left figure) and P11 (right figure).  The solid diamonds reflect the $a_\mathrm{P}$ from \citet{niet05a}, the solid squares are the calculated points for $a_\mathrm{Pc}$, and the ``X's'' are calculated points for dates ten days from the date of the $a_\mathrm{P}$. }
\label{fig:1}
\end{figure*}

The components of $ z_\mathrm{c}$ values are listed in Table~\ref{tab:2}.  Figure~\ref{fig:1} plots the $a_\mathrm{P}$ and $a_\mathrm{Pc}$ values versus $D$ for each spacecraft.  The correlation coefficient between the $a_\mathrm{P}$ and $a_\mathrm{Pc}$ is 0.85 for all data points.  Without the P11 80/66 data point, which is the most uncertain measurement \citep{niet05a}, the $a_\mathrm{P}$ and $a_\mathrm{Pc}$ correlation coefficient is 0.93.

\begin{table}
\caption{The values of the constants of Eq.~(\ref{eq:26}). }
\label{tab:1}
\begin{tabular}{lll}
\hline
\hline
{Parameter}
&{value}
&units\\ 
\hline
$K_\mathrm{dp}$&$\phantom{-}1.30 \times 10^{-28}$& erg$^{-1}$~AU$^{-2}$\\
$K_\mathrm{p}$&$\phantom{-}2.10 \times 10^{-26}$& erg$^{-1}$~AU$^{-1}$\\
$K_\mathrm{f}$&$\phantom{-}5.38 \times 10^{-22}$& erg$^{-1}$\\
$K_\mathrm{co}$&$-5.98 \times 10^{6}$& erg~AU$^{-1}$\\
\hline\\
\end{tabular}
\end{table}

\begin{table}
\scriptsize
\caption{The components of Eq.~(\ref{eq:26}) and $A_\mathrm{sun}$ for each data point. }
\label{tab:2}
\begin{tabular}{lrrrrrr}
\hline
\hline
{Craft Date}
&{$D$\phantom{00}}
&{$ K_\mathrm{dp} D P $}
&{$ K_\mathrm{p} P $}
&{$ K_\mathrm{f} F $}
&{$X$\phantom{0}}
&$A_\mathrm{sun}$\\ 
&(AU)&$\times 10^{-14}$&$\times 10^{-14}$&$\times 10^{-14}$&$\times 10^{-13}$&($^{\circ}$)\\
\hline
P10 82/19&25.80&-1.83&-5.74&15.40&0.78&124\\
P10 82/347&27.93&-2.06&-5.97&16.77&0.87&165\\
P10 83/338&30.68&-2.43&-6.42&18.54&0.97&167\\
P10 84/338&33.36&-2.82&-6.85&20.27&1.06&175\\
P10 85/138&36.57&-4.43&-9.81&21.17&0.69&12\\
P10 86/6&36.53&-3.34&-7.42&21.12&1.04&144\\
P10 87/80&40.92&-4.38&-8.66&23.96&1.09&70\\
P10 88/68&43.34&-4.76&-8.90&26.69&1.30&82\\
P10 89/42&45.37&-5.04&-8.99&26.82&1.28&109\\
P11 77/270&6.49&-0.24&-3.05&2.99&-0.03&43\\
P11 80/66&8.42&-0.26&-2.49&4.23&0.15&166\\
P11 82/190&11.80&-0.49&-3.34&6.41&0.26&109\\
P11 83/159&13.13&-0.55&-3.42&7.25&0.33&148\\
P11 84/254&17.13&-0.99&-4.68&9.83&0.42&71\\
P11 85/207&18.36&-1.01&-4.47&10.62&0.51&121\\
P11 86/344&23.19&-2.09&-7.31&13.74&0.43&16\\
P11 87/135&22.41&-1.40&-5.06&13.23&0.68&151\\
P11 88/256&26.62&-2.01&-6.12&15.93&0.78&88\\
P11 89/316&30.29&-2.85&-7.62&18.29&0.78&36\\
\hline
\end{tabular}
\end{table}

The error bars in Fig.~\ref{fig:1} reflect the uncertainty from Table II of\citet{ande02} except for P11 80/66 where the uncertainty is from\citet{niet05a}.  The stochastic variable of the unmodeled acceleration was sampled in ten-day or longer batches of data \citep{ande02}.  Starting at the P11 80/66 data point, the average extended over many months \citep{niet05a}.  The ``X's'' in Fig.~\ref{fig:1} plot the calculated data point for a ten-day change from the date of the $a_\mathrm{P}$.  Some data points showed little change between the two dates of calculation.  Others showed moderate change.  Because the value of $a_\mathrm{Pc}$ depends on the closeness of matter to the LOS, a change over ten days is due to a body close to the LOS and to long integration times.  Therefore, the closeness of matter to the LOS introduces an uncertainty for even ten-day integration times that was unaccounted in the error budget.

The $a_\mathrm{Pc}$ calculation reproduces the subtler effects of the PA noted by \citet{ande02}.

\subsection{\label{sec:annual}Annual periodicity}

\begin{figure*}
\begin{minipage}[c]{0.45\textwidth}
\includegraphics[width=\textwidth]{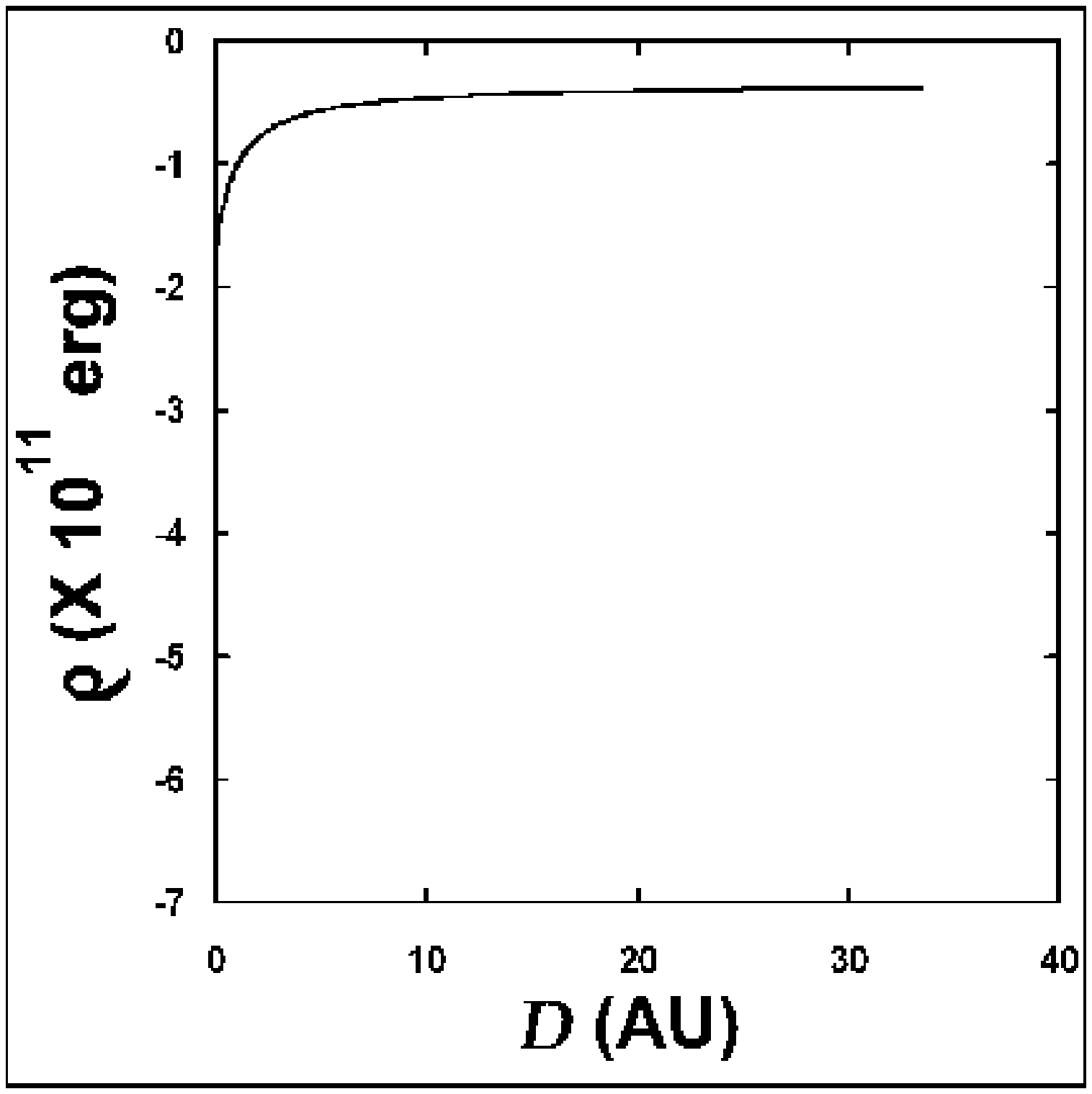}
\end{minipage}
\begin{minipage}[c]{0.45\textwidth}
\includegraphics[width=\textwidth]{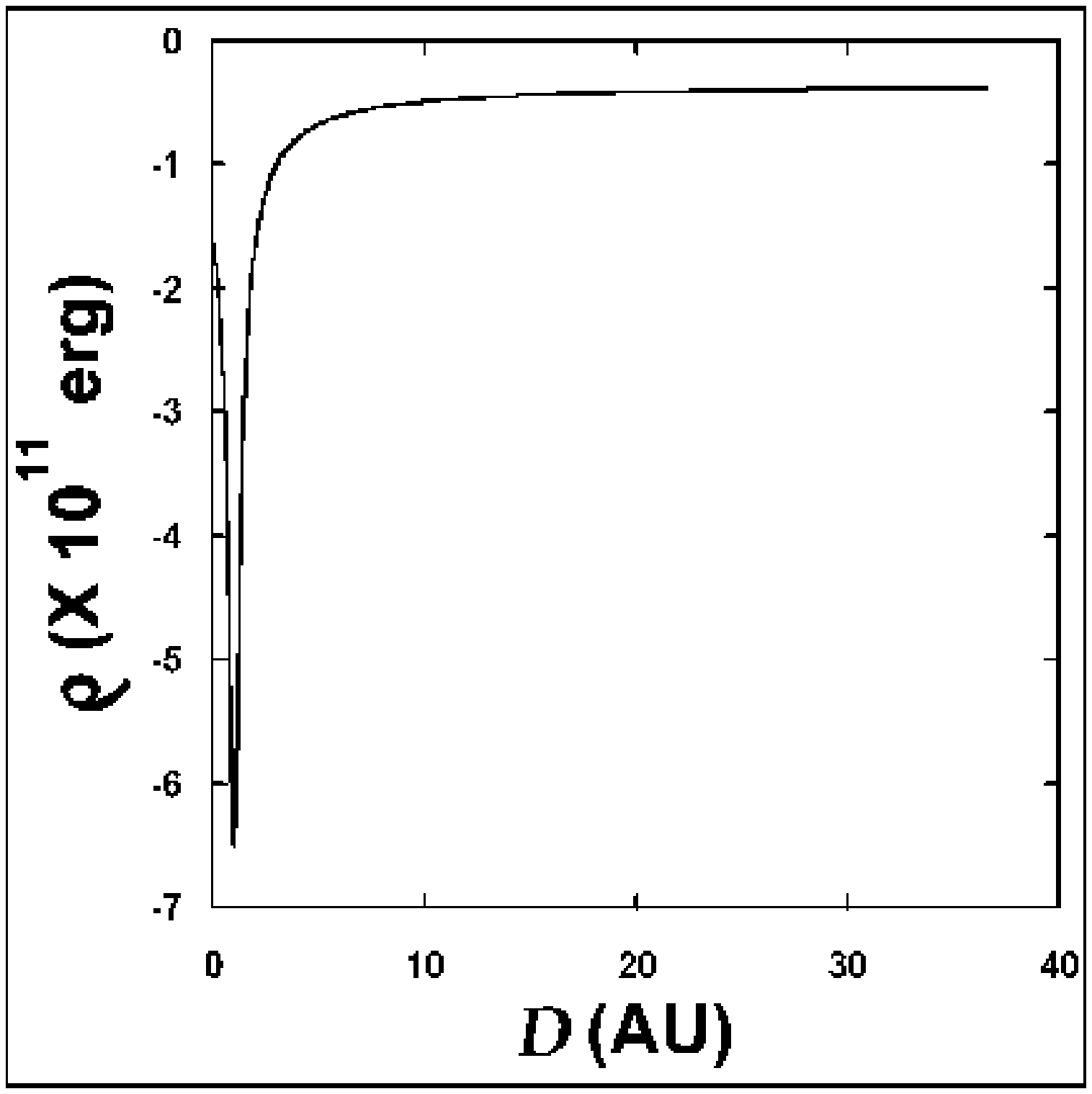}
\end{minipage}
\caption{Plots of the $\rho$ versus geocentric distance $D$ along the LOS.  Data is for P10 on 84/338 (left figure) $a_\mathrm{P} =8.43  \times 10^{-8}$ cm s$^{-2}$ and for P10 on 85/138 (right figure) \mbox{$a_\mathrm{P} =7.67 \times 10^{-8}$ cm s$^{-2}$}.  The $A_\mathrm{sun} \approx 175^\circ$ on P10 84/338 and $A_\mathrm{sun} \approx 16^\circ$ on P10 85/138.}
\label{fig:2}
\end{figure*}

Figures~\ref{fig:2} show the $\rho$ value along the LOS versus $D$ on dates when the angle $A_\mathrm{sun}$ (degrees) between the LOS and a line from the Earth to the Sun was $<60^\circ$ and when $A_\mathrm{sun}> 120^\circ$.  

Table~\ref{tab:2} lists the $A_\mathrm{sun}$ for each data point.  On the dates that $A_\mathrm{sun} < 60^\circ$, the $a_\mathrm{P}$ and $a_\mathrm{Pc}$ were considerably lower.  The correlation coefficient between the $a_\mathrm{P}$ and $a_\mathrm{Pc}$ without P10 85/138 ($A_\mathrm{sun} \approx 12^\circ$), P11 86/344 ($A_\mathrm{sun} \approx 16^\circ$), and P11 80/66 (Saturn encounter) is 0.97.  The low $A_\mathrm{sun}$ value combined with long integration time causes larger uncertainty.

To test the effect of the Sun on $a_\mathrm{Pc}$, the calculation was repeated with the Sun excluded ($a_\mathrm{Psun}$).  Figure~\ref{fig:3} shows the $a_\mathrm{Pc}$ values versus $D$ for the spacecraft from Fig.~\ref{fig:1} and $a_\mathrm{Psun}$.  The effect of the Sun is to cause the annual variation of the $a_\mathrm{Pc}$.  Therefore, the cause of the PA is also the cause of the annual periodicity.  

\subsection{Difference of $a_\mathrm{p}$ between the spacecraft}

\begin{figure}
\includegraphics[width=0.5\textwidth]{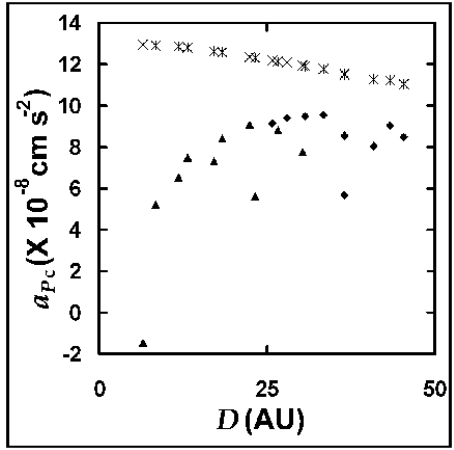}
\caption{ Plots of the $a_\mathrm{Pc}$ data versus geocentric distance $D$ for P10 (solid diamonds) and P11 (solid triangles).  The stars plot the data points for $a_\mathrm{Pc}$ with the Sun excluded ($ a_\mathrm{Psun}$). }
\label{fig:3}
\end{figure}

The P10 data included one of nine dates ($\approx 0.11$) with $A_\mathrm{sun}<60^\circ$ and five of nine dates ($\approx 0.56$) with $A_\mathrm{sun}>120^\circ$.  The P11 data included three of ten dates ($\approx 0.30$) with $A_\mathrm{sun}<60^\circ$ and four of ten dates ($\approx 0.40$) with $A_\mathrm{sun}>120^\circ$.  Figure~\ref{fig:3} shows the trend of $a_\mathrm{Psun}$ versus $D$ between P10 and P11 data points.  At $D>10$ AU, the $a_\mathrm{Psun}$ appears to be a linear function of $D$.  At $D<10$ AU, the Sun's influence is to lower $a_\mathrm{P11}$ more than $a_\mathrm{P10}$.  

\begin{figure*}
\begin{minipage}[c]{0.45\textwidth}
\centering \includegraphics[width=\textwidth]{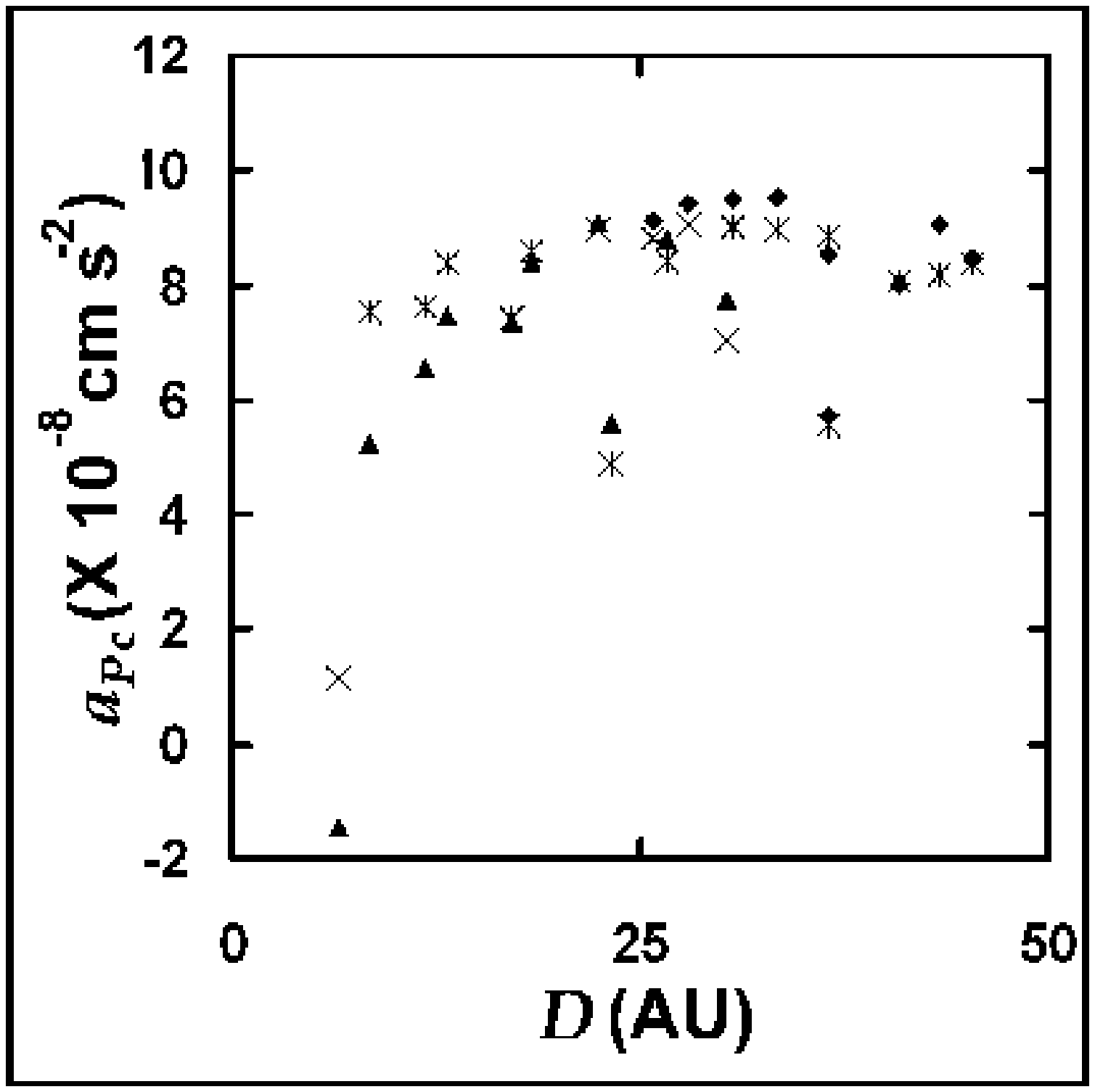}
\end{minipage}
\begin{minipage}[c]{0.45\textwidth}
\centering \includegraphics[width=\textwidth]{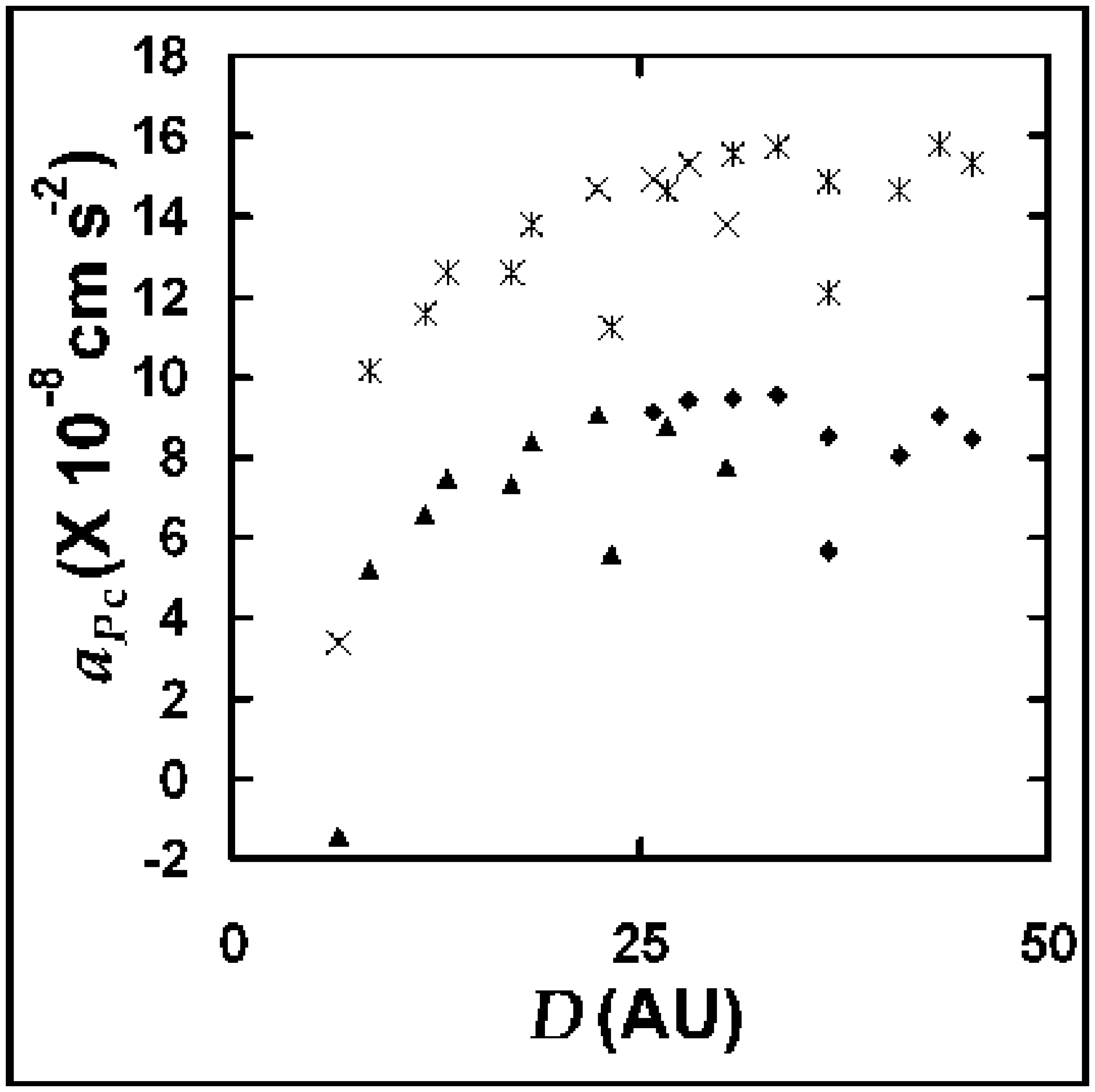}
\end{minipage}
\caption{Plots the $a_\mathrm{Pc}$ versus geocentric distance $D$ for P10 (solid diamonds) and for P11 (solid triangles).  The data points of $a_\mathrm{Pc}$ excluding the outer planets $a_\mathrm{Pplanets}$ (left), and the $a_\mathrm{Pc}$ excluding the galaxy mass $a_\mathrm{Pgal}$ (right) are shown as stars.}
\label{fig:4}
\end{figure*}

The SPM also suggests the mass of the planets and the mass of the galaxy has an influence on the $\rho$ field.  Figure~\ref{fig:4} plots the $a_\mathrm{Pc}$ for the spacecraft, the $a_\mathrm{Pc}$ excluding the outer planets ($a_\mathrm{Pplanets}$), and the $a_\mathrm{Pc}$ excluding the galaxy mass ($a_\mathrm{Pgal}$) versus $D$.

Because the outer planets are opposite the Sun for P10, the effect of the planets on $a_\mathrm{Pc}$ of P10 is less than P11.  However, as $D$ of P11 increases, the $a_\mathrm{Pplanets} \longrightarrow a_\mathrm{Pc}$.

From the galaxy scale perspective, the spacecraft in the solar system appears as near the large mass of the Sun and inner planets.  The effect of the galaxy mass appears to decrease the $a_\mathrm{Pc}$ nearly uniformly for P11.  The outer P10 data points show a trend to increasing relative effect of the galaxy mass.  The orbit of P10 is closer to the Sun-Sgr.A* axis than P11 and the $D$ of P10 is greater than the $D$ of P11.  However, this effect is within the uncertainty level.  

The difference in $a_\mathrm{P10}$ and $a_\mathrm{P11}$ noted by\citet{ande02} results primarily from the effect of the Sun.  A secondary effect is the effect caused by increasing $D$ and a small effect of the planets on P11 which declines as $D$ increases.  The SPM expects the galaxy mass to have a small difference between $a_\mathrm{P10}$ and $a_\mathrm{P11}$ caused by their different travel directions.  The $a_\mathrm{P}$ is not constant as the CHASMP software assumed \citep{ande02}.  Therefore, the varying $a_\mathrm{P}$ may explain the difference between the {\textit{Sigma}} and CHASMP program methods for P10 (I) and their closer agreement at P10 (III) \citep[Table I]{ande02}.

\subsection{Slow decline in $a_\mathrm{P}$}

The plot in Fig.~\ref{fig:3} suggests d$a_\mathrm{Psun}$/d$D$ at $D<10$ AU is nearly zero, followed by a decline and then a flattening.  

The radio signal measurements are from and to Earth.  At small $D$, the relative effect of the Sun-Earth distance is larger than at farther $D$.  As $D$ increases the Solar System effect appears to approach a single mass located at the barycenter of the Solar System.  Therefore, $a_\mathrm{P}$ declines and approaches a constant value dictated by $\rho \propto -R^{-1}$.  However, the SPM expects that at much greater $D$, the effect of the galaxy mass will increase to cause a difference in the $a_\mathrm{P}$ values between the spacecraft.

\subsection{Saturn encounter}

\begin{figure}
\includegraphics[width=0.5\textwidth]{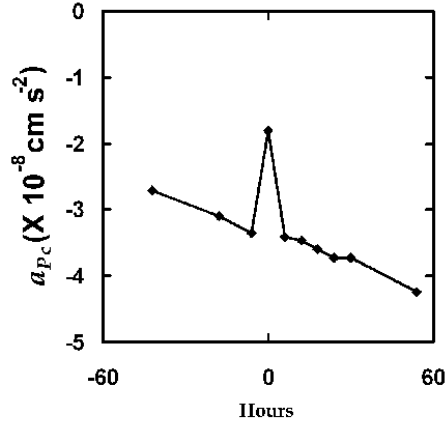}
\caption{Plot of $a_\mathrm{Pc}$ versus the hours from the closest approach of P11 to Saturn.}
\label{fig:6}
\end{figure}

Figure~\ref{fig:6} shows a plot of $a_\mathrm{Pc}$ versus the hours from the closest approach of P11 to Saturn on P11 79/244 ($A_\mathrm{sun} \approx 8^\circ$).  The plot shows the $a_\mathrm{Pc}$ varies widely over a period of hours.  The negative $a_\mathrm{Pc}$ is a redshift.  As seen in Fig.~\ref{fig:1}, the SPM is consistent with the P11 77//270 ($A_\mathrm{sun} \approx 43^\circ$) data point at the beginning of the Saturn encounter of a near zero blueshift.  

\subsection{Large uncertainty of P11 80/66}

Because the P11 80/66 ($A_\mathrm{sun} \approx 166^\circ$) data point extends over a relatively long time, the rapidly varying $a_\mathrm{Pc}$ seen in Fig.~\ref{fig:6} is consistent with the uncertainty in the P11 80/66 data point. 

The $a_\mathrm{Psun}$ data points for P11 77/270 and P11 80/66 in Fig.~\ref{fig:3} have only a slightly lower slope than the later data points.  The planets gravity well is in a larger gravity well of the Sun, which is in an even larger galaxy gravity well.  The change from the Sun $\rho$ versus $D$ curve to a planet gravity well causes a smaller $K_\mathrm{f} F$ term relative to $K_\mathrm{p} P$.  Table~\ref{tab:2} lists $|K_\mathrm{f} F| < |K_\mathrm{p} P| $ for the P11 77/270, where ``$| \, |$'' means ``absolute value'', and $|K_\mathrm{f} F| > |K_\mathrm{p} P| $ for other data points.  Without the Sun gravity well in the calculations, $|K_\mathrm{f} F| > |K_\mathrm{p} P| $ for all data points.  Therefore, the $a_\mathrm{Psun}$ for the P11 77/270 data point is consistent with the other data points.

\subsection{Cosmological connection}

\begin{figure}
\includegraphics[width=0.5\textwidth]{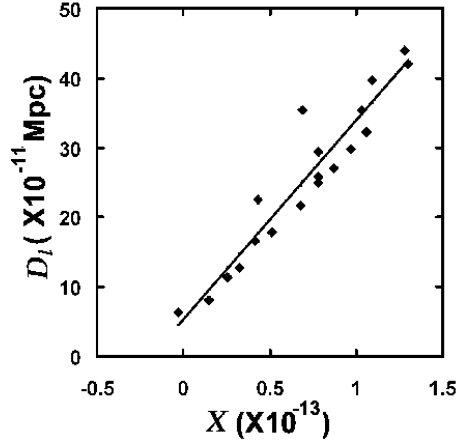}
\caption{Plot of the light travel distance $D_\mathrm{l} (\times 10^{-11}$ Mpc) versus $X (\times 10^{-13}$) for both spacecraft.  The line is a plot of $ D_\mathrm{l}= 2800 X + 5 \times 10^{-11}$ Mpc.}
\label{fig:5}
\end{figure}

The SPM obtains the $H_\mathrm{o}$ value by $z \longrightarrow \exp( -X ) \,-1 \approx \, -X$, where $X<0$ in \citet{hodg} and $X>0$ in this paper.  Figure~\ref{fig:5} is a plot of $D_\mathrm{l}$ with the units changed to Mpc versus $X$.  The straight line is a plot of the least-squares fit of the data.  The line is 
\begin{eqnarray}
D_\mathrm{l}&=& (2800 \pm 200 \mathrm{Mpc} ) X + (5 \pm 2) \times 10^{-11} \mathrm{Mpc} \nonumber \\*
&\approx& -\frac{c}{H_\mathrm{o}} z
\label{eq:28b}
\end{eqnarray}
at 1$\sigma$ and with a correlation coefficient of 0.95, where \mbox{$ H_\mathrm{o} =106 \pm 8$~km~s$^{-1}$~Mpc$^{-1}$}.

The PA and the $z$ of cosmology are the result of the same $\rho$ effect on light.  In the cosmological $z$ calculation, the $z$ follows the Hubble law if $\rho \propto R^{-1}$.  In the gravity well, $z$ follows the negative Hubble law if $\rho \propto -R^{-1}$.  The presence of other galaxies near the path of the light causes $P$ and $F$ variation of $z$.  This is also the effect of matter close to the LOS in the PA.  In the SPM, the Hubble law and $a_\mathrm{P} \approx cH_\mathrm{o}$ are manifestations of the Newtonian spherical property. 

\section{\label{sec:disc}Discussion and conclusion}

The variation of $a_\mathrm{p}$ through the sidereal day remains to be investigated.  The specific location on Earth relative to the Earth's center and the timing of the transmitting and receiving signals are required.  The observation that the period is a sidereal day rather than a solar day suggests the SPM will be consistent with this observation, also.

The $K$ values were calculated using data obtained in ``interesting'' environments.  This means the $\rho$ field was changing rapidly and long integration times were used.  The $K$ values could be calculated more accurately with longer integration time and in environments with little change in the $\rho$ field.  That is, the $K$ values could be calculated more accurately in ``uninterseting'' environments when the Sun and planets are farther from the LOS or by using shorter integration times.

The $D$ at which the $a_\mathrm{Pgal}$ becomes noticeable is curiously close to the orbit of Pluto.

Near planetary bodies the deep gravitational potential well causes a net frequency change of such a low value as to be undetectable.  Because the mass of the spacecraft is much smaller then planetary bodies, the very small potential well of the spacecraft allows an observable frequency shift.  

This SPM of the PA is that the frequency shift is always present and was present at less then 10 AU.  However, as noted in the P11 77/270 data point and in the Saturn encounter calculation (Fig.~\ref{fig:6}), the value of $a_\mathrm{p}$ changes and may be a redshift.  Therefore, further analysis of the spacecraft data can be a test of this model with additional caveats that a shorter integration period be used, that $A_\mathrm{sun} \approx 180^{\circ} \pm 30^{\circ}$, and that the $a_\mathrm{p}$ be considered a variable (the {\textit{Sigma}} software).  The Jupiter to Saturn travel period with a minimal radial movement would reduce the Doppler effect.  Because this model is a non-Doppler frequency shift model, the verification of this effect has significant cosmological implications.  

The SPM suggests gravity is a manifestation of a curved $\rho$ field.  Sources, Sinks, and matter produce the curvature.  In GR, matter curves Space and curved Space\footnote{The uppercase ``S'' on Space indicates the view of GR and the $\rho$ field of the SPM.  A lower case ``s'' indicates the neutral backdrop, Euclidean space.} constrains the dynamics of matter.  In the SPM, the $\rho$ field acts like the Space of GR and space is a neutral backdrop (flat) in which distance is calculated using timing events such as using Cepheid stars.  Therefore, the proposition that matter warps the $\rho$ field is reasonable.  Space in GR and the $\rho$ field in the SPM is curved, ubiquitous, and corporeal.  Therefore, the objection to Newtonian mechanics of ``action at a distance'' is removed.  Calling the stuff of the $\rho$ field ``Space'' as GR does is tempting but confusing when dealing with the Euclidean, neutral backdrop type of space. 

\citet{hodg} suggests the flow of the $\rho$ field from Sources to Sinks causes the cell structure of galaxy clusters.  An analogy is fluid flow from Sources to Sinks causes a Rankine oval.  The B band luminosity $L$ of a galaxies indicated the effect of a Source or Sink on the intergalactic redshift.  In the redshift calculation, the use of $L$ may also be the net effect of the Sources and Sinks and of the matter around galaxies.  A problem with this interpretation is that at the outer part of the galaxy, the gravitational mass of the galaxy is still larger than the $\rho$ field effect as evidenced by the net attraction of hydrogen.  Also, $L$ was found to be proportional to spiral galaxy rotation curve parameters \citep{hodg2} and to galaxy central parameters \citep{hodg3} that do not include the total mass of the galaxy.  The paradox between the two uses of $L$ may be resolved if the effect of galaxy matter on intergalactic scales is zero.  In fluid flow, an object placed in an otherwise unconstrained flow produces local turbulence downstream.  However, eventually the flow of the fluid appears as if the object was absent.  This is consistent with the presence of high metallicity matter in the intergalactic medium (e.g. \citealt{arac}).  Therefore, the proposition that matter is neither a Source nor a Sink is consistent with the cluster and galaxy observations.  Further, if the $\rho$ field is flat, $\vec{\nabla} \rho =0$ and there is no flow.  In such a condition such as between galaxy clusters, the static field of matter extends throughout the flat field.

The observations that confirm GR also confirm the $\rho$ field effect because each has the same effect on matter.  Gravitational lensing is the effect of the $\rho$ field on photons perpendicular to their travel path.  The effect herein is the effect of the $\rho$ field on photons along their travel path.  When observations consider matter that moves through space such as $z$ measurements and gravitational lensing observations, the amount, curvature, and flow of the $\rho$ field must be considered. 

I speculate GR corresponds to the SPM in the limit in which the Sources and Sinks may be replaced by a flat and static $\rho$ field such as between cluster cells and on the Solar System scale at a relatively large distance from a Source or Sink.  On the galaxy and galaxy cluster scale the $\rho$ field is significantly curved from the proximity of Sources and Sinks and GR fails \citep{sell}.

For a sample of 19 data points with published Pioneer Anomaly ``acceleration'' values, the SPM was found to be consistent with the observation of not only the value of the Pioneer Anomaly, but also with the subtler effects noted in \citet{ande02}.  The SPM is consistent with the general value of $a_\mathrm{P}$; with the annual periodicity; with the differing $a_\mathrm{P}$ between the spacecraft; with the discrepancy between {\textit{Sigma}} and CHASMP programs at P10 (I) and their closer agreement at P10 (III); with the slowly declining $a_\mathrm{P}$; with the low value of $a_\mathrm{P}$ immediately before the P11's Saturn encounter; with the high uncertainty in the value of $a_\mathrm{P}$ obtained during and after the P11's Saturn encounter; and with the cosmological connection suggested by $a_\mathrm{P} \approx cH_\mathrm{o}$.  The SPM has outstanding correlation to observed data when the long integration time combined with rapidly changing $\rho$ field is insignificant.  Because the gradient of the $\rho$ field produces a force on matter, the effect of the $\rho$ field warp appears as the curvature of space proposed by GR that causes the gravitational potential of matter.  

\section*{Acknowledgments}

The anonymous reviewer of \citet{hodg} inspired this investigation of the Pioneer Anomaly within the context of the SPM.

I acknowledge and appreciate the financial support of Maynard Clark, Apollo Beach, Florida, while I was working on this project.





\end{document}